\documentstyle[preprint,revtex]{aps}

\begin{document}

\begin{title}
Replica Real Space Renormalization
Group for Spin Glasses
\end{title}

\bigskip

\author{J.R.L. de Almeida and S. Coutinho}

\begin{instit}
Departamento de F\'{\i}sica
Universidade Federal de Pernambuco
50670-901 Recife, PE, Brasil
\end{instit}

\vspace{1.5cm}

\centerline{\bf ABSTRACT}

\bigskip

We construct a real space renormalization group (RG) approach
for Ising spin glasses  on hypercubic lattices within the scheme
of the Migdal-Kadanoff approximation using replicas. Our replica
symmetric solution yields results consistent with simple
decimation previously obtained and the introduction of breaking
of replica symmetry within the RG is discussed, which inserts in
a natural fashion non-linear RG into the problem.

\vspace{1.5cm}


\newpage

\noindent
{\bf 1. Introduction}

\bigskip

The study of the magnetic properties of the so-called spin
glasses (random magnetic systems with quenched disorder and
competing interactions) has been challenging the imagination of
many workers in the field, and up to now the only fairly
complete theory is the one associated with the solution of the
mean field Sherrington-Kirkpatrick (SK) model (for an excellent
account of the development of the field see Fisher and Hertz
1991), which suffers from the original sin of having size
dependent and infinite range interactions. Nevertheless, these
limitations of the model might be forgiven when one considers
bona fide  models and from them obtain the known mean field
theories either for the pure ferromagnetic case (Kac 1968) or
spin glasses (de Almeida 1992).

Usually the mean field results hold at very large
dimensionalities, give the qualitative aspects that a theory in
finite dimensions should encompass, and are obtained without
much effort. Spin glasses changed this altogether for even its
mean field theory (assumed as the solution of the SK model) took
many years to be worked out, and only in the past few years a
phase transition in Ising $3d$ systems has been largely accepted
(see Fisher and Hertz 1991).

To study the nature of the phase transitions in 3d systems a
natural framework would be to employ the renormalization group
techniques (see Wallace and Zia 1978 for a review). However,
from the beginning the absence of a correct picture of the
nature of the cooperative fluctuations in spin glasses will be
hanging over any results which might be obtained.

Among the various forms of implementing the renormalization
group approach, real space methods in lattice models has been
widely used for its conceptual and technical simplicity, in
particular the Migdal-Kadanoff renormalization group (MKRG)
approximation.  In this work we study Ising spin glasses on hypercubic
lattices using the MKRG approximation in conjuction with the
replica method. The study of Ising spin glasses using the MKRG
approximation is not new (Ney-Nifle and Hilholst 1993 and
references therein). What is new in our work, we believe, is the
introduction of the replica method and new parameters allowing
breaking of replica symmetry. Thus a full consistence between the
mean field (SK) theory and a renormalization group approach may
emerge.

\vspace{1.5cm}

\noindent
{\bf 2. The Model and its RG}

\bigskip

Let us consider a model described by the following Hamiltonian
\begin{equation}
{\cal H} = - \sum_{(i,j)} J_{ij} \sigma_{i} \sigma_{j}
\hspace{0.5cm} , \hspace{0.5cm} \sigma_{i} = \pm 1
\end{equation}

\noindent
where $(i,j)$ labels the pair of nearest neighbor sites on a
hypercubic lattice with random exchange interactions coupling
constants $J_{ij}$'s assumed to have zero mean and variance $J$
of a gaussian distribution. The bond average of the replicated
partition function of the model
(1) is (Edwards and Anderson 1975)
\begin{equation}
\left< {\cal Z}^{n} \right> = Tr \; \exp \left\{ \frac{K^{2}}{2}
\sum_{\alpha , \beta} \sum_{(i,j)} \sigma^{\alpha}_{i}
\sigma^{\alpha}_{j} \sigma^{\beta}_{i} \sigma^{\beta}_{j} \right\}
\end{equation}

\noindent
where $< \cdots >$ means bond average, $K = \beta J$, and
$\alpha , \beta = 1,2,...n$ are replica indices. Let us now
perform an approximate RG transformation on (2) within the
Migdal-Kadanoff approach. Here we carry this out by working with
the effective Hamiltonian in replica space  as defined by (2):
\begin{equation}
{\cal H}_{eff} = \frac{K^{2}}{2} \sum_{\alpha , \beta}
\sum_{(i,j)} \sigma^{\alpha}_{i} \sigma^{\alpha}_{j}
\sigma^{\beta}_{i} \sigma^{\beta}_{j}
\end{equation}

\noindent
As far as we know, none of the existing works on spin glasses
using the RG approach used (3) as the starting point. First,
consider the simple decimation scheme using a rescaling factor
$b=2$, and a new rescaled lattice and interactions are obtained
from the previous ones using the clusters indicated in figure 1.
Calling $S_{i}$ and $S_{j}$ the spins of the rescaled lattice
each connection in the cluster shown in figure 1  will produces a
contribution to the effective interaction between $S_{i}$ and $S_{j}$,
which is obtained
by tracing out the replicated intermediary ($\sigma^{\alpha}$)
spins in the following expression
\begin{equation}
I_{i,j} = Tr_{\{ \sigma \}} exp \left\{ \frac{K^{2}}{2}
\sum_{\alpha , \beta} \left( S^{\alpha}_{i} \sigma^{\alpha}
S^{\beta}_{i} \sigma^{\beta} + S^{\alpha}_{j} \sigma^{\alpha}
S^{\beta}_{j} \sigma^{\beta} \right) \right\}
\end{equation}

\noindent
Using standard transformation equation (4) assumes the form
\begin{equation}
I_{i,j} = \int Dx_{1} \int Dy_{1} exp \left\{ K_{1}
(x_{1},y_{1}) \sum_{\alpha} S^{\alpha}_{i} S^{\alpha}_{j} \right\}
\end{equation}

\noindent
where
\begin{equation}
K_{1} (x_{1},y_{1}) = tgh^{-1}  \left[ tgh (K x_{1}) tgh (K
y_{1}) \right]
\end{equation}

\noindent
and
\begin{equation}
Dx_{i} = \frac{dx_{i}}{\sqrt{2 \pi}} exp \left\{ -
\frac{x^{2}_{i}}{2} \right\}
\end{equation}

\noindent
{}From (5) it is seen that each connection in figure 1 contributes with a
term $K_{1}(x_{1},y_{1})$ to the effective interaction between
$S^{\alpha}_{i}$ and $S^{\alpha}_{j}$, all replicas with the
same weight, each term with a probability distribution specified
by (7). The new effective interaction is then given by
\begin{equation}
K'_{ij} = \sum^{2^{d-1}}_{\ell = 1} K_{\ell} (x_{\ell}, y_{\ell})
\end{equation}

\noindent
for d--dimensional hypercubic lattices. The
probability distribution of the interactions in the decimated
lattice is then given by
\begin{equation}
P(K') = \int  \delta \left[ K' - \sum^{2^{d-1}}_{\ell =1}
tgh^{-1} \left( tgh (Kx_{\ell})tgh (Ky_{\ell}) \right) \right]
\prod^{2^{d-1}}_{\ell =1 } Dx_{\ell } Dy_{\ell}
\end{equation}

\noindent
whose mean deviation $<K'^{2}>$ is
\begin{eqnarray}
<K'^{2}> & = & \frac{1}{2 \pi} \int dz \int dK' \int_{\ell} \prod
Dx_{\ell} Dy_{\ell} (K')^{2} exp \left\{ iz K' - \right.
\nonumber \\
& - & \left. i z \sum_{\ell} tgh^{-1} \left[ tgh (kx_{\ell}) tgh
(Ky_{\ell}) \right] \right\}
\end{eqnarray}

\noindent
which may be recast into the form ($b=2$)
\begin{equation}
<K'^{2}> = 2^{d-1} \int \int Dx \; Dy \left[ tgh^{-1} \left( tgh
(Kx) tgh (Ky) \right) \right]^{2}
\end{equation}

\noindent
Analogously, one obtains for the fourth moment of the distribution
\begin{equation}
<K'^{4}> = 3 \left( <K'^{2}> \right)^{2} + 2^{d-1} \left[
<u^{4}> - 3 <u^{2}>^{2} \right]
\end{equation}

\noindent
where $<u^{2}>$ is the double integral in equation (11) and
\begin{equation}
<u^{4}> = \int Dx \int Dy \left[ tgh^{-1} \left( tgh (Kx) tgh
(Ky) \right) \right]^{4}
\end{equation}

\noindent
Equations (11) and (12) could have been obtained without the
need of introducing the replicas shown in equation (4). We
exhibit them to stress its replica symmetric nature. They are
the result of a one step decimation starting from a gaussian
bond distribution and the last term in (12) is a measure of the
deviation from a gaussian of the new iterated distribution. They
may be extended to provide an iterative map between the moments
of the effective bond distributions at each RG transformation.
Here we take the approximation of considering (11) and (12)
valid at each RG step which amounts to approximating the
iterated probability distributions by gaussians, which is a good
approximation as $d$ assume large values.

{}From (11) we obtain that there is a phase transition only for
$d>2$, and $T_{c} \simeq 0.95$, 2.12, 3.51, 5.32 for $d=3$,4,5, and 6,
respectively. The correlation coefficient $\rho =
<K'^{4}>/3(<K'^{2}>)^{2}$, which is one for a gaussian
distribution, takes the values at $T_{c}$, $\rho = 1.11$ ($d=3$);
1.11 ($d=4$); 1.08 ($d=5$) 1.05 ($d=6$) justifying the above
mentioned approximation. The correlation-length exponent ($\nu$)
defined by the flow away from the non-trivial fixed point of
(11) is given by
\begin{equation}
\left. \frac{d<K'^{2}>^{1/2}}{dK} \right|_{T_{c}} = 2^{1/2 \nu}
\end{equation}

\noindent
and we find $\nu = 1.41$ ($d=3$), $\nu = 0.77$ ($d=4$), $\nu =
0.60$ ($d=5$), $\nu =  0.55$ ($d=6$). For comparison the high
temperature series expansion for these exponents (Klein et al
1991) are $\nu = 1.37$ ($d=3$), $\nu = 0.95$ ($d=4$), $\nu = 0.73$
($d=5$) for a $\pm J$ distribution. Others critical indices may
be obtained in the usual way invoking their relationship
obtained through usual scaling.

It might be worth mentioning that phenomenological scaling (Bray
and Moore 1985, Bray 1988 and references therein) claim that
the ordered state is governed by a zero-temperature fixed point,
such that the width of the renormalized probability distribution
at low temperatures has a growth characterized by an exponent
$y$, i.e., $K' \sim 2^{y}K$. From (11) we find numerically that
$y \simeq (d-5/2)/2$, yielding a lower critical dimension $d_{c}
\approx 2.5$. For comparison, numerical studies (Bray and Moore
1985, Bray 1988) estimate $y \simeq - 0.30$ ($d=2$) and
$\simeq 0.2$ ($d=3$) while our result gives $y \simeq - 0.25$
($d=2$) and $y \simeq 0.25$ ($d=3$). However, we note that our
results thus far have been obtained within the context of
replica symmetric approach. An improved treatment may upset the
above flow of the renormalized probability distribution. While
equation (11) gives a stable zero-temperature fixed point for $d
> d_{c}$, a proper account of frustration may turn it unstable
as in highly frustrated systems studied by McKay et al (1982).

We consider now the question of introducing breaking of replica
symmetry within the present approach. The renormalization group
framework for studing critical properties can rarely be carried
out exactly. In general, one has to resort to some
approximations and frequently introduce some parameters to
optimize the results (see Wallace and Zia 1978). In the case
under study, we may take this step writing equation (4) in the
form
\begin{equation}
I_{i,j} = Tr_{\{ \sigma \}} exp \left\{ \frac{K^{2}}{2}
\sum_{\alpha , \beta} q_{\alpha \beta} \left[ S^{\alpha}_{i}
\sigma^{\alpha} S^{\beta}_{i} \sigma^{\beta} + S^{\alpha}_{j}
\sigma^{\alpha} S^{\beta}_{j} \sigma^{\beta} \right] \right\}
\end{equation}

\noindent
where the $q_{\alpha \beta}$'s are new parameters introduced
into the theory to be determined say, for instance,
variationally from a free energy. Breaking of replica symmetry (RSB)
within the real space RG may then be pursued by iterating (15).
A first step RSB scheme (see below) leads to a iterative map
involving several parameters which is under study.

In this work we have only analysed the replica symmetric case
$q_{\alpha \beta} = q = 1$ and leave for future work other
cases.

\vspace{1.5cm}

\noindent
{\bf 3. Discussion}

\bigskip

In this work we have reported, we believe, the outline of the
renormalization group in real space as should be applied to spin
glasses (here only Ising systems were considered but extension
to other systems is straightforward). Only the naive replica
symmetric case was considered and results consistent with
previous works were obtained (Southern and Young 1977, Curado
and Meunier 1988, Bray 1988, Ney-Nifle and Hilhorst 1993, and
references therein). It was our aim in this work to explicitly
point out the replica symmetric (RS) nature of the results,
obtain only the simplest results within the RS solution, and to
lay the ground for extension of the theory. We are not aware of
any similar discussion except in a work of one of the present
authors (de Almeida 1993), where it was argued that it is in the
nature of these complex systems (spin glasses) that they require
the use of nonlinear renormalization group in their analysis.
This may completely change standard universality rules as
obtained for uniform systems (critical exponents depending only
on the space dimension and on the number of spin components)
which have been confirmed numerically through Monte Carlo
calculation (Bernardi and Campbell 1993) for $d=3$ and 4, using
several distinct forms for the probability distribution of the
interactions. Moreover, a study of the multifractal spectra of
the order parameter for short range Ising spin glasses in $d=3$
(Coutinho et al 1993) indicate a nontrivial $f( \alpha
)$-function for $T \leq T_{c}$ thus suggesting the complex
nature of a proper scaling for spin glasses. In the present
approach, nonlinear renormalization group sets in imediately as
soon as one start breaking replica symmetry among the $q_{
\alpha \beta}$ in equation (15), in a fashion similar to the
breakdown of linear response theory and the Fisher relation in
the spin glass phase (Bray and Moore 1980).

In the last few years many works either experimental (Lefloch et
al 1993 and references therein) or numerical (Badoni et al 1993,
Hetzel et al 1993, and references therein) have argued in favour
of the mean field SK picture in finite dimensions. There is even
the suggestion that for Ising spin glasses there is no
qualitative difference in their behaviour for all dimensions
above the lower critical one (Bhatt et al 1991). Our preliminar
analysis of breaking replica symmetry using equation (15)
following a Blandin's like seminal scheme (Blandin 1978),
indicates $y$ exponents much larger than in the RS case
suggesting the possibility of existence of an de
Almeida-Thouless transition line (de Almeida and Thouless 1978)
within the present context. In addition, RSB introduces an
unfolding of the bonds probability distribution much like in
line with the RSB picture yielding a probability law for the
probability law of the overlap of the order parameters
(M\`{e}zard et al 1984). All these topics deserve further
investigation which we leave for future work. In this letter we
hope to have added another step forward in the long and hard
spin glass way.

\bigskip

\noindent
{\bf Acknowledgments}

We should like to thank CNPq, FINEP and CAPES (Brazilian
government granting agencies) and that of FACEPE
(Pernambuco state granting agency) under the grant APQ 211--1.05/91
for partial financial support. One of us (JRLA) thanks
I.A. Campbell for very useful correspondence and preprints.

\newpage

\noindent
{\bf References}

\bigskip

\noindent
\begin{description}
\item[ ] Badoni D, Ciria J C, Parisi G, Pech J, Ritort F and
Ruiz J J 1993 Europhys. Lett. \underline{21}, 495.
\item[ ] Bernardi L and Campbell I A preprints (1993).
\item[ ] Bhatt R N, Reger J D and Young A P 1991 J. Appl. Phys.
\underline{69}, 5219.
\item[ ] Blandin A 1978 J. Physique \underline{C6}, 1568.
\item[ ] Bray A J and Moore M A 1980 J. Phys. C: Solid State
Phys. \underline{13}, 419 \\
$\overline{\hspace{2.0cm}}$ 1985 J. Phys. C: Solid State Phys.
\underline{18}, L927.
\item[ ] Bray A J 1988 Comment. Condens. Mat. Phys.
\underline{14}, 21
\item[ ] Coutinho S, de Almeida J R L and Curado E M F 1994,
Proc. of the IFIP Second International Working Conference on Fractals
in the Natural and Applied Sciences---FRACTAL 93, Ed. M M Novak, Elsevier, to
appear.
\item[ ] Curado E M F and Meunier J L 1988 Physica
\underline{A149}, 164.
\item[ ] Edwards S F and Anderson P W 1975 J. Phys. F: Metal
Phys. \underline{5}, 965.
\item[ ] de Almeida J R L and Thouless D J 1978 J. Phys. A:
Math. Gen \underline{11}, 983.
\item[ ] de Almeida J R L 1991 J. Phys. A: Math. Gen.\\
$\overline{\hspace{2.0cm}}$ 1993 J. Phys. A: Math. Gen.
\underline{26}, 193.
\item[ ] Fisher K H and Hertz J A 1991 Spin Glasses (Cambridge:
Cambridge Univ. Press).
\item[ ] Hetzel R E, Bhatt R N and Singh R R P 1993 Europhys.
Lett. \underline{22}, 383; Kac M 1968 Statistical Physics, Phase
Transitions and Superfluidity (Brandeis University) vol.1
(New York: Gordon and Breach).
\item[ ] Klein L, Adler J, Aharony A, Harris A B and Meir Y 1991
Phys. Rev. \underline{B43}, 11249.
\item[ ] Lefloch F, Hamman J, Ocio M and Vincent E 1993
Europhys. Lett. \underline{18}, 647.
\item[ ] McKay S R, Berker A N and Kirkpatrick S 1982 Phys. Rev.
Lett. \underline{48}, 767.
\item[ ] M\`{e}zard M, Parisi G, Sourlas N, Toulouse G and
Virasoro M 1984 J. Physique \underline{45}, 843.
\item[ ] Ney-Nifle M and Hilhorst H J 1993 Physica
\underline{A193}, 48.
\item[ ] Sherrington D and Kirkpatrick S 1975 Phys. Rev. Lett.
\underline{35}, 1972.
\item[ ] Southern B W and Young A P 1977 J. Phys. C: Solid State
Phys. \underline{10}, 2179.
\item[ ] Wallace D J and Zia R K P 1978 Rep. Prog. Phys.
\underline{41}, 1.
\end{description}

{\bf Figure Captions}
\begin{description}
\item[Figure 1:] Migdal--Kadanoff $b=2$ cluster. There are $ p=2^{d-1}$
connections
joining the $ S_{i}'s$ spins in d-dimensions.
\end{description}
\end{document}